\documentclass[prb,preprint]{revtex4-1}

\usepackage{amsmath}  
\usepackage{amsfonts} 
\usepackage[pdftex]{graphicx,color} 
\usepackage{hyperref} 

\hypersetup{pdfstartview={XYZ null null 0.99}, bookmarksnumbered=true, bookmarksopen=true, colorlinks=true, linkcolor=black, citecolor=black, urlcolor=black, filecolor=black, anchorcolor=black, runcolor=black}

\begin{document}

\title{Analyzing the structure of basic quantum knowledge for instruction}

\author{Giacomo Zuccarini}
\email{giacomo.zuccarini@uniud.it}
\affiliation{University of Udine, Department of Mathematics, Computer Science and Physics, via delle scienze 206, 33100, Udine, Italy}

\date{\today}

\begin{abstract}
In order to support students in the development of expertise in quantum mechanics, we asked which concepts and structures can act as organizing principles of the non-relativistic theory. The research question has been addressed in a multi-step process based on the analysis of categorization studies, on a content analysis of a sample of undergraduate textbooks and on the results of existing research on learning difficulties. The answer consists in seven concept maps, intended as models of the organizing principles of quantum knowledge needed to account for the results of measurement and time evolution. By means of these instruments, it is possible to visualize and explore the different facets of the interplay between the vector structure of the quantum states and the operator structure of the observables, and to highlight the educational significance of the relations between observables. The maps can be used by instructors as a support for helping students build a well-organized knowledge structure and by researchers as a basis for the design of investigations into student understanding. While this framework may be adapted to different approaches and interpretive stances, it provides indications in favor of a spin-first approach over a waves-first one.
\end{abstract}

\maketitle

\section{Introduction}

Twenty years of educational research show that the development of expertise in basic, non-relativistic quantum mechanics (QM) is a challenging task for students.\cite{Marshman2015b} They adopt survival strategies for performing reasonably well during their course, becoming proficient at solving algorithmic problems, but they have difficulty mastering concepts and applying the formalism to answer qualitative questions related to quantum concepts and processes.\cite{Singh2008} In several respects, upper-level undergraduate students taking QM display patterns of difficulties analogous to those reported for introductory students, such as a lack of a robust knowledge structure; in both cases, the consistency of their reasoning strongly depends on the context.\cite{Marshman2015b} Results of researches conducted at the end of the course \cite{Singh2001} and at the beginning of graduate instruction\cite{Singh2008} suggest that knowledge fragmentation - i.e. the organization of knowledge in small, disconnected pieces, which can be productively applied locally but which lack global consistency - often persists after a whole semester or year of intensive instruction on the topic. The importance of helping students overcome fragmentation and develop expertise in basic QM at undergraduate level is further supported by two additional facts. First, only a fraction of students goes on to graduate physics programs in QM: for those who do not, the undergraduate course will be their last quantum exposure. Secondly, there is a growing body of evidence to suggest that students' understanding of undergraduate QM is not substantially improved by instruction at a graduate level.\cite{Carr2009, Michelini2014, Emigh2015, Passante2015b}

Another important issue, specifically concerning the teaching of QM, is related to the
organization of the subject-matter content for instruction. As testified by empirical research\cite{Lin2010} and by recent reviews of quantum textbooks,\cite{Harshman2019} there is still no agreement among faculty members as to which approach is preferable to introduce QM at undergraduate level: waves-first or spin-first? The proponents of the spin-first approach believe that it provides a simple two-dimensional vector space to teach the foundations of QM whereas those who discuss, for instance, the infinite square well, first believe that spin is too abstract and continuity with the content discussed in the earlier courses is important.\cite{Lin2010}

Research has shown that expert knowledge in physics is hierarchical in structure (major principles and concepts at the top, facts and equations occupying the lower levels), highly organized (each node is richly linked to other relevant knowledge within the hierarchy) and conditionalized for efficient use: that is, it cannot be reduced to sets of isolated propositions, because procedures for applying principles and concepts are bundled with them, as are the contexts in which those principles/concepts apply.\cite{Bransford2000, Mestre2001} As a consequence, choices made on the organization of the subject-matter content as well as on its elementary features necessarily influence the expertise development process. However, educational investigations on the content structure of QM are currently lacking, and though some of the textbooks attempt to locally organize the information by giving a summary at the end of each single chapter,\cite{McIntyre2012} they do not include, so far as we know, comprehensive attempts to organize the information at a global level.

In this paper, we endeavour to analyze the structure of basic quantum knowledge, in order to identify what entities should be at the top of the knowledge pyramid, as well as their interconnections and the related procedures that make possible an efficient use of this knowledge, such as applying the formal tools of the theory in order to make qualitative predictions on quantum systems. In the rest of the paper, we shall call this specific knowledge network ``the organizing principles'' of the subject-matter. This analysis aims to provide a basis for helping students overcome fragmentation and build a well-organized knowledge structure in QM, as well as for giving indications on which
organization of the content is preferable from an educational standpoint.
The research question we endeavour to answer is the following:

\begin{itemize}
    \item [RQ1:] which concepts and structures can act as organizing principles in basic QM?
\end{itemize}

\section{Instruments and methods}
The methods used to conduct the analysis are drawn from categorization studies and the Model of Educational Reconstruction (MER), a theoretical framework designed for converting domain specific knowledge into knowledge for instruction.\cite{Duit2012}

Categorizing or grouping together problems based upon similarity of solution has been employed both in introductory physics research \cite{Chi1981a} and in upper-level QM \cite{Lin2010} to infer how physics knowledge is structured in memory and to assess physics students' level of expertise. In classical mechanics, differences between novice and experienced problem solvers have been identified both in the way they organize their knowledge of physics and how they approach problems. Novice problem solvers use literal objects from the surface attributes of the problems as category criteria (such as ``spring'' problems or ``inclined plane'' problems), whereas experienced problem solvers consider the physics concept or principle used to solve the problem when deciding on problem categories, such as grouping conservation of energy problems together.\cite{Docktor2014} For the purpose of identifying the organizing principles of quantum knowledge at upper-undergraduate level, we referred to the categorization study by S.-Y. Lin and C. Singh,\cite{Lin2010} the only one available on the subject.

Another useful strategy to navigate this search is provided by the MER. According to it, a fundamental stage in converting domain specific knowledge into  knowledge for instruction is called \textit{elementarization}, i.e. the identification of the entities within a complex content domain which may be viewed as elementary features (e.g., basic phenomena, basic principles and general laws), the combination of which helps explain the different aspects of the scientific content considered. Research methods mainly draw on qualitative content analysis of leading textbooks and key publications on the topic under inspection, to be conducted from the standpoint of science education. This analysis can benefit significantly from research into student ideas on the specific subject at hand, as student conceptions may provide a new view of science content and hence allow another, deeper, understanding of the subject under examination. After examining the categorization study, we drew on content analysis of a sample of textbooks in the search of the elementary features needed to explain our findings. The results obtained were tested and further processed by means of existing research on student understanding of QM.

\section{The two processes}

In the categorization study by S.-Y. Lin and C. Singh,\cite{Lin2010} 22 physics juniors and seniors in two undergraduate
quantum mechanics courses and six faculties were asked to categorize 20 QM problems based upon similarity of solution, using one or more categories for each problem. The goal of the study was to investigate differences in categorization between faculties and students and whether there are major differences in the ways in which individuals in each group categorize QM exercises. Three of the six faculties involved were asked to evaluate the quality of each category, scoring it as `good' (which is assigned a score of 2), `moderate' (1) or `poor' (0), without using identifiers and with the categorizations by the faculties and students jumbled up. In several respects, this categorization task gave different results from those obtained in studies on introductory physics. However, categorizations by faculty members were rated higher overall than those by students.

For this reason, we assumed that categories with the best score (5-6) reflect a highly organized knowledge of the subject, and can be used as instruments to identify central features of basic QM.
Consequently, instead of focusing on the differences in categorization between groups and individuals in each group (as the authors did), we chose to focus on the best scoring categories, grouping them by similarity of topic and counting the number of problems they were placed in. The categories \textit{time dependence of expectation value}, \textit{time evolution of wavefunction}, \textit{stationary state}, are all specific instances of the quantum evolution of systems in the absence of measurement, and therefore were collected under the label \textit{Time Evolution}. The categories \textit{measurement}, \textit{expectation-value}, \textit{collapsed wavefunction}, \textit{expectation-value and uncertainty} concern different aspects of the peculiar behavior of systems in measurement, and therefore were collected under the label \textit{Measurement}. The categories \textit{expansion in eigenfunction} and \textit{Fourier transform} refer to mathematical procedures used to represent a state vector with respect to another eigenbasis, and were collected under the label \textit{Change of Basis}. The remaining categories, i.e. \textit{symmetry argument} and \textit{spin}, were considered separately. Table \ref{TAB:1} displays the number of problems in which each collective category was placed. A pattern emerges: while in introductory physics categorization is typically based on the fundamental principles used in order to solve problems (e.g., conservation principles and Newton's laws), in QM it is primarily based on the process involved (either measurement or time evolution in the absence of measurement) and secondarily on a mathematical procedure often needed to extract from the state information on measurement/time evolution (change of basis). As the authors of the study observe: ``faculty members noted that the fundamental principles, e.g. conservation laws, are also important in understanding quantum processes but they are not the focus of an upper-level undergraduate quantum mechanics course.''
\begin{table}[h!]
\centering
\caption{Grouping the best scoring categories in the study by S.-Y. Lin and C. Singh. \cite{Lin2010}}
\begin{ruledtabular}
\begin{tabular}{l c p{3cm}}
Category & Instances  \\
\hline	
Measurement & 10 \\
Time-Evolution & 9  \\
Change of Basis & 4 \\
Spin & 1 \\
Symmetry Argument & 1 \\
\end{tabular}
\end{ruledtabular}
\label{TAB:1}
\end{table}

Further evidence supporting the central role of the two processes in education is provided by the most comprehensive review of student difficulties at upper-undergraduate level published up to date.\cite{Singh2015b} If we read this review looking for issues concerning these processes, we notice that - while many topics are discussed - difficulties with measurement and time evolution, and related procedures, occupy the bulk of the study's report: about 8/13 pages dedicated to the presentation of reasoning difficulties, organized into two sections for time evolution (\textit{Difficulties with the time dependence of a wave function}, 1,5 pages, and \textit{Difficulties with the time dependence of expectation values}, 1,5 pages) and two on measurement (\textit{Difficulties with measurements and expectation values}, 4 pages, and \textit{Difficulties involving the uncertainty principle}, 0,5 pages), while the section devoted to difficulties with Dirac notation has mostly to do with measurement (0,5 pages).

As concerns the nature of measurement and time evolution, we observe that, if we accept the projection postulate, both of them represent fundamental evolution processes. Independently from this choice, measurement remains at the core of quantum theory, as the evolution in time concerns the probability distributions of measurement outcomes.

Differently from categorization studies in Newtonian mechanics, our search for the organizing principles of quantum knowledge does not end here. As a matter of fact, the two processes under consideration do not represent the concepts needed to solve quantum problems, but rather the processes which \textit{need} to be explained. In conclusion, in order to identify the nodes at the top of the pyramid of quantum knowledge, their interconnections and the attached procedures, we have to determine what kind of network might account for the possible results of measurement and time evolution, both at a qualitative and quantitative level (RQ1.1). At the same time, as the complexity of knowledge structure in QM is due to both the requisite conceptual and mathematical knowledge,\cite{Lin2010} this network should also explain how `information regarding the two processes'\footnote{The expression refers to the data required to make predictions on the result of the processes.} is encoded in the formal representations of quantum systems (RQ1.2).  The two sub-questions are a refinement of the tasks required to answer RQ1.

\section{The organizing principles of basic quantum knowledge}

In order to find candidate answers to the sub-questions formulated at the end of the last section, we examined how the two processes are presented and discussed in a sample of textbooks written for upper-undergraduate courses. In particular, we searched for the ways in which relevant concepts and procedures are interconnected, with the goal of identifying knowledge structures that are able to give a generally valid and full account of the results of the processes in non-relativistic QM. For this purpose, we selected four textbooks, two of them starting with QM in one spatial dimension, and two adopting a spin-first approach. As to the former, we analyzed Griffiths' QM text\cite{Griffiths2018} which, according to a recent survey on the use of textbooks in upper-level QM courses,\cite{Dubson2009} currently dominates the market. Alongside Griffiths', we also included Gasiorowicz's textbook\cite{Gasiorowicz2007}, quite popular among faculty members when they were taught QM as undergraduate students.\cite{Dubson2009} As concerns spin-first, we selected McIntyre's text\cite{McIntyre2012} - to our knowledge, the only QM textbook with contributions from physics education researchers - and Townsend's,\cite{Townsend2012} one of the first undergraduate textbooks adopting this approach.

Here we present the general picture we obtain in the light of the content of the textbooks and the results of research on learning difficulties. Our answer to RQ1.1 is summarized in seven concept maps, intended as models of the organizing principles of quantum knowledge needed to account for the results of measurement and time evolution both at a qualitative and quantitative level. In these maps, we represent general pathways of solutions which start from the required task, i.e. making predictions on the result of a process: measurement of an observable on a state, time evolution of a state, time evolution of the distribution of an observable. Next, we inspect the entities under examination and apply the procedures and the criteria needed to solve the task. The pathway ends with the outcome of the process, first at a qualitative level, then at a quantitative one. A pathway of solution is considered of qualitative nature when it does not involve complex quantitative procedures (e.g. the change of basis).

\begin{figure}[!ht]
    \centering
       \fbox{\includegraphics[width=15cm]{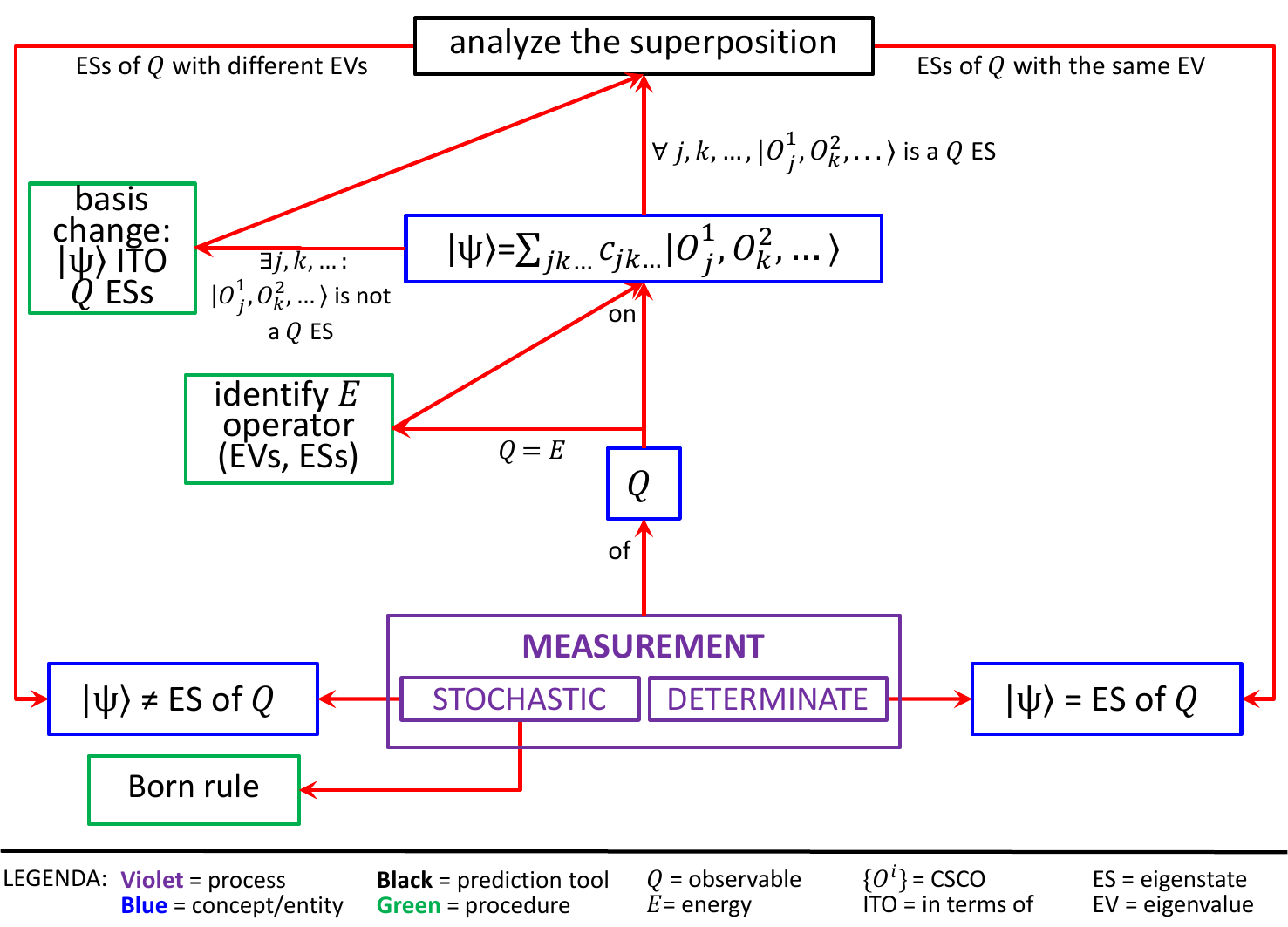}}
    \caption{Measurement on a superposition state.}\label{FIG:3}
\end{figure}

Three maps directly emerge from the content analysis. While the textbooks widely differ in how and when the concepts (e.g., eigenstates), the formal structures (e.g., operators) and the contexts (e.g., spin-1/2 particles in a magnetic field or harmonic oscillators) are introduced and discussed, the methods they use to find the results of the processes are the essentially the same. As concerns the measurement of an observable $Q$ on a state $|\psi \rangle$ (see fig. \ref{FIG:3}), the generally valid sequence is the following: a) $|\psi \rangle$ is represented as a superposition - trivial or not - of simultaneous eigenstates of a complete set of commuting observables (from now on: CSCO, indicated on the maps as $\{O^i\}$). This is always the case, although only Tonwsend and Gasiorowicz introduce and discuss the concept of CSCO; b) if necessary, a change of basis is performed to represent $|\psi \rangle$ as a superposition of $Q$ eigenstates; c) the analysis of the superposition allows to predict whether, at a qualitative level, the result will be determinate or stochastic; d) in the latter case, quantitative results are obtained by applying the Born rule.
If $Q$ represents energy, we need to know in advance its eigenvalues and eigenstates, and thus the solution to the eigenvalue problem.

Alongside the aforementioned sequence, all of the textbooks present other ways to give information on the results of measurement, i.e. the uncertainty relation $\Delta A \Delta B\geq \frac{1}{2}|\langle [ \hat{A},\hat{B} ] \rangle|$, exploiting the symmetry in the potential, so as to find the expectation value of observables on energy eigenstates (see, e.g., Griffiths, pp.50-51), and the formula $\langle\psi|\hat{Q}|\psi\rangle$. However, in all cases the information we get is partial (constraints on the results, the expectation value), and can be obtained only in a limited number of cases, where these methods are applicable or appropriate. As a consequence, while they offer convenient shortcuts in given contexts and should be taught in an undergraduate course, they are not included among the organizing principles.

\begin{figure}[!ht]
    \centering
       \fbox{\includegraphics[width=15cm]{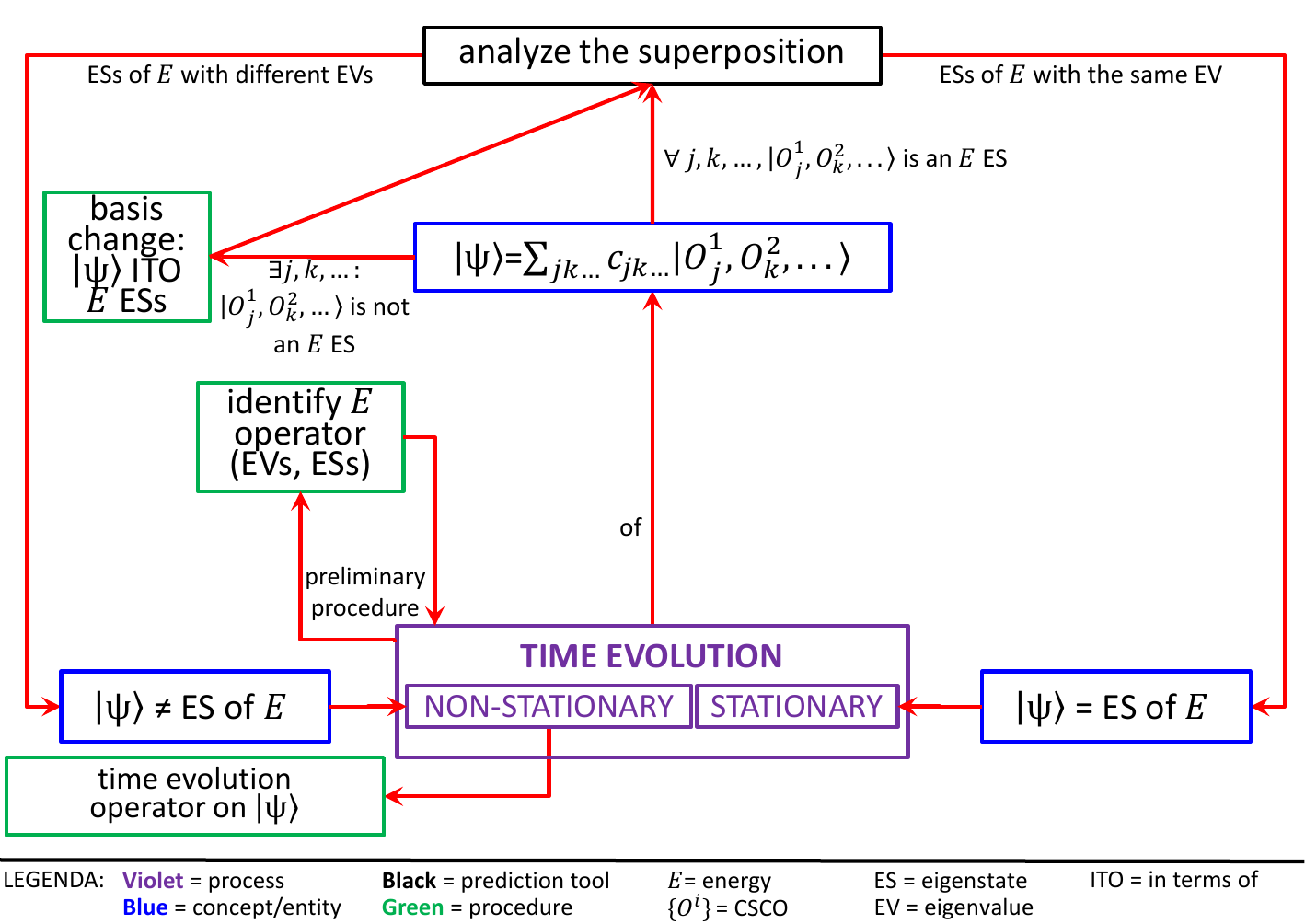}}
    \caption{Time evolution of a superposition state.}\label{FIG:4}
\end{figure}

As concerns the time evolution of a state $|\psi\rangle$, the textbooks in our sample rely almost exclusively on the following sequence: a) $|\psi\rangle$ is a superposition of simultaneous eigenstates of a CSCO; b) if necessary, a change of basis is performed to represent the state in terms of $E$ eigenstates; c) the superposition is analyzed to determine whether the state is stationary or not; d) in the latter case, quantitative results are found by applying to $|\psi\rangle$ the time evolution operator (see fig. \ref{FIG:4}).

Two of the textbooks in our sample also present an alternative approach: the Feynman propagator method. However, only Townsend devotes one chapter to the path integrals, while Griffiths mentions them in passing. As Townsend remarks, in non-relativistic QM, the main utility of this approach is ``in the alternative way it gives us of viewing time evolution'', and NOT in predicting its results, i.e. ``in explicitly determining the transition amplitude''  (see Townsend, pp. 290-291), inasmuch as it ``is not especially practical in most problems'' (ibid. p. 300). As a consequence, propagators are not included in the maps.

\begin{figure}[!ht]
    \centering
       \fbox{\includegraphics[width=14cm]{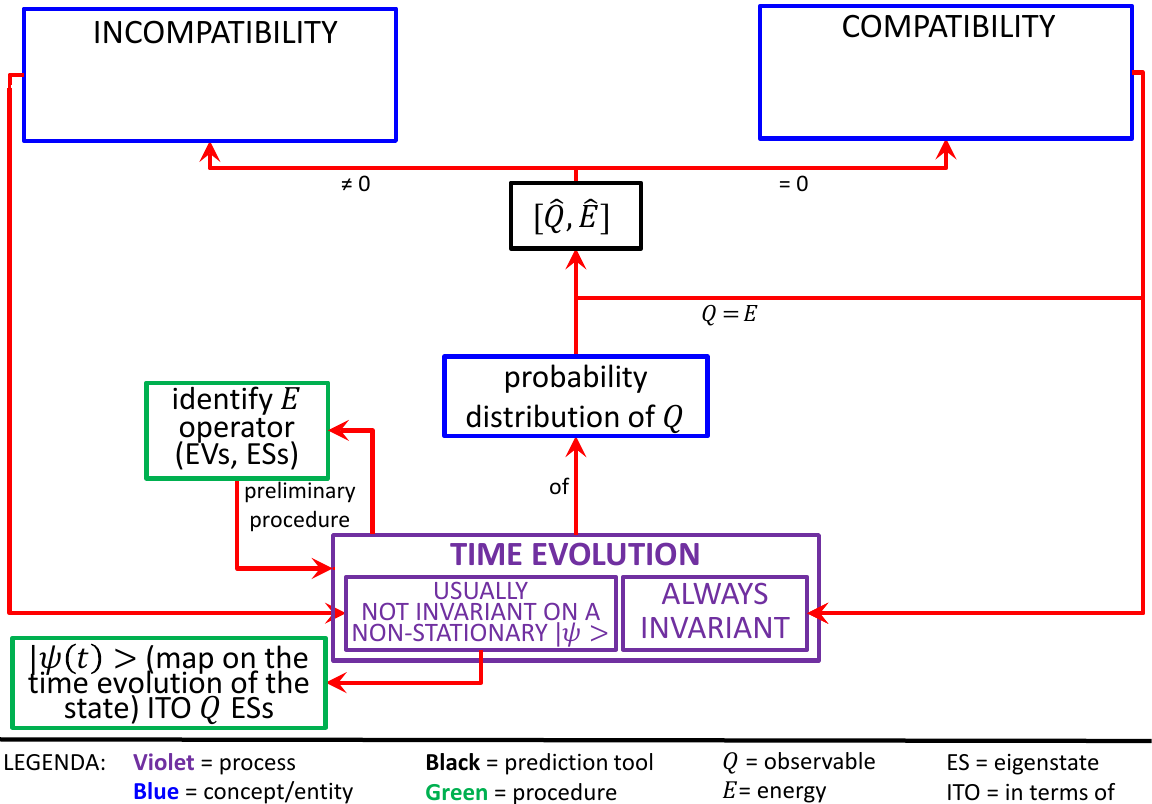}}
    \caption{Time evolution of the probability distribution of an observable.}\label{FIG:5}
\end{figure}

The third map addresses the evolution of the probability distribution of an observable $Q$, and is merely a visual representation of what is implied by
$[\hat{Q} , \hat{E}]$ (see fig. \ref{FIG:5}). If the observables are incompatible, the time evolution of the distribution of $Q$ on a non-stationary state $|\psi \rangle$ is obtained by determining $|\psi(t)\rangle$ (map on the time evolution of the state) and its representation in terms of $Q$ eigenstates. The prediction tool is the analysis of the commutator.

\begin{figure}[!ht]
    \centering
       \fbox{\includegraphics[width=10cm]{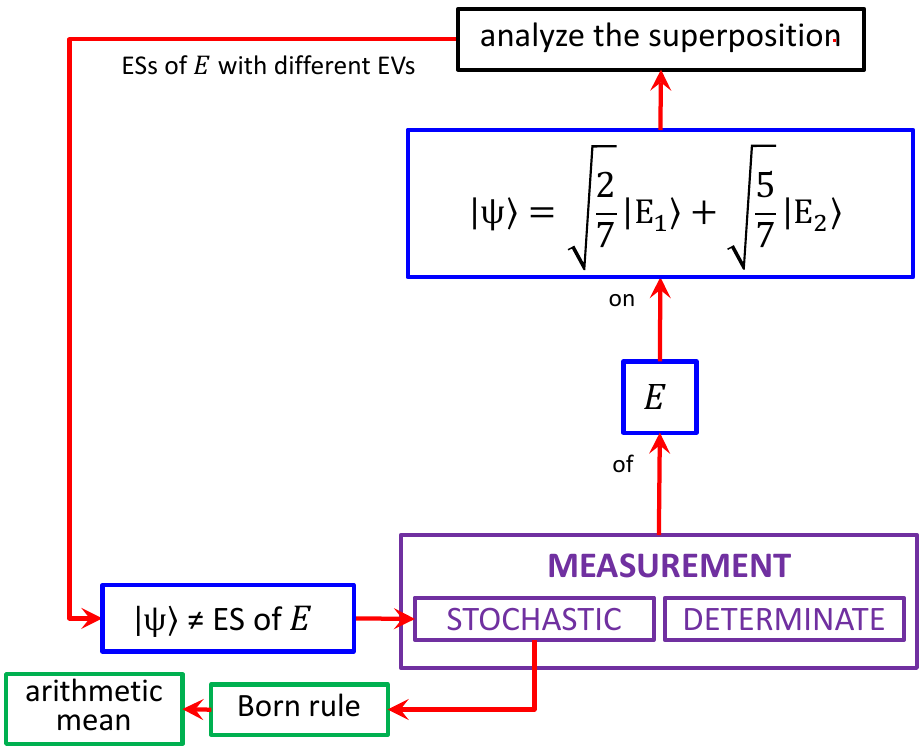}}
    \caption{Question: write down the probability of measuring an energy value and the expectation value.\cite{Singh2008}}\label{FIG:1}
\end{figure}

\begin{figure}\centering
\begin{tabular}{|r|l|} \hline
    \includegraphics[width=.5\textwidth]{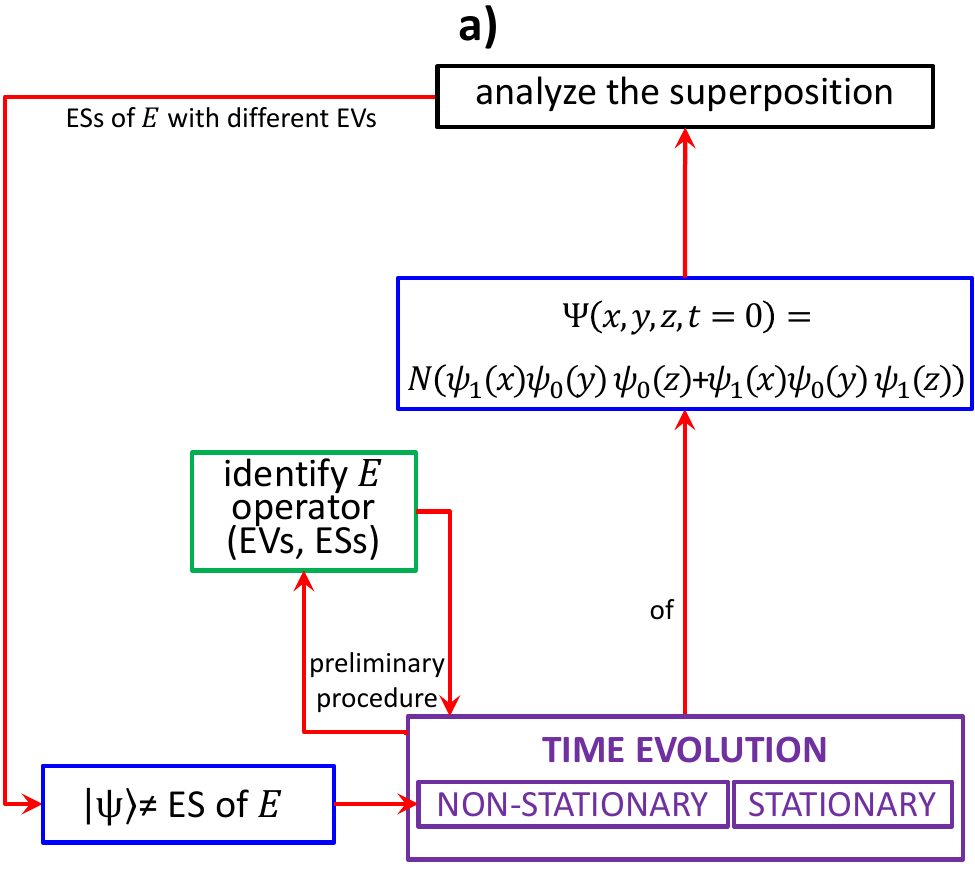} &
    \includegraphics[width=.5\textwidth]{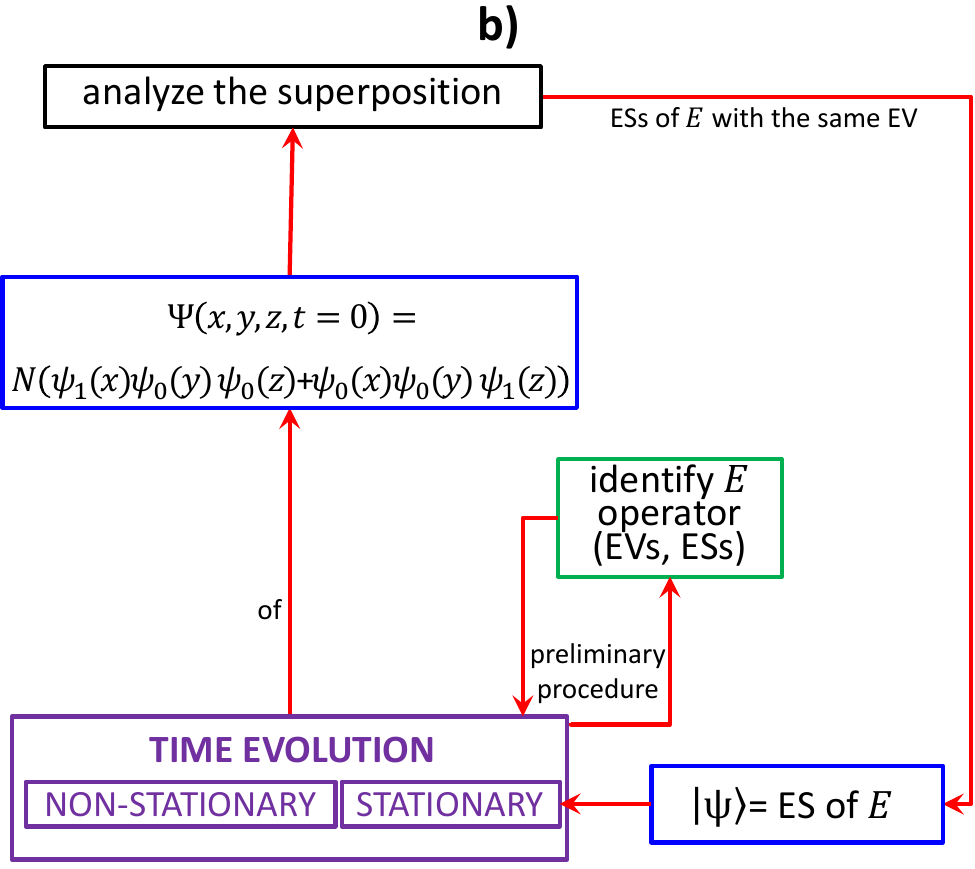} \\ \hline
\end{tabular}
\caption{Question: consider the following wave functions for a particle in a three-dimensional isotropic harmonic oscillator potential. Does the state change as time evolves? Explain.\cite{Emigh2015}}\label{FIG:2}
\end{figure}

Fig. \ref{FIG:1} and \ref{FIG:2} show how the knowledge structure represented in the first two maps is applied to find qualitative pathways of solution to survey questions administered in educational studies.

While the pathways displayed in the third map are quite straightforward, the first two general maps present various elements of interest.

First, by looking at these maps, we notice a strong structural similarity as concerns the knowledge needed for finding the results of measurement and of the time evolution of the state. In particular, in both maps, the tool for predicting the outcome of the processes at a qualitative level is the analysis of the superposition.

Secondly, knowledge about the specific context of the problem (e.g., harmonic oscillator, hydrogen atom, etc...) comes into play whenever energy is involved and, at least partially, when we perform a change of basis. In the transition from classical mechanics to QM, the context of a problem acquires a greater importance, as a change of context affects the measurable values of energy, its eigenstates, the commutation relations of the Hamiltonian, and may involve tensor products of different Hilbert spaces. The greater significance of contexts in QM was also stressed by faculties interviewed by S.-Y. Lin and C. Singh.\cite{Lin2010}

Thirdly, in the maps, we considered a superposition of eigenstates of a CSCO with discrete spectra. If this is not the case, the sum is replaced by an integral. However, this  change does not affect the structure of the pathways for finding the results of the processes.

Finally, despite the general validity of the first two maps, from an educational standpoint they are incomplete, inasmuch as they do not include qualitative pathways of solution to very simple tasks, e.g. deciding whether a position eigenstate is a stationary state or not, which is exactly the kind of knowledge students struggle to acquire.\cite{Singh2015b} Research has shown that deep and unexpected learning difficulties emerge in the interplay between state vectors and operators in the Hilbert space. Singh and Marshman report that a very common difficulty is assuming that eigenstates of operators corresponding to all physical observables are the same. More in general, their review includes whole subsections discussing \textit{Difficulties in distinguishing between eigenstates of operators corresponding to different observables} (concerning measurement) and \textit{Difficulties in distinguishing between stationary states and eigenstates of operators corresponding to observables other than energy}. The textbooks offer very few examples on how to use the commutation relations between observables - and therefore knowledge about their systems of eigenstates - in order to find the results of measurement and the time evolution of the state, while our first two maps incorporate this kind of knowledge only implicitly in the procedure of the change of basis.

\begin{figure}[!ht]
    \centering
       \fbox{\includegraphics[width=15cm]{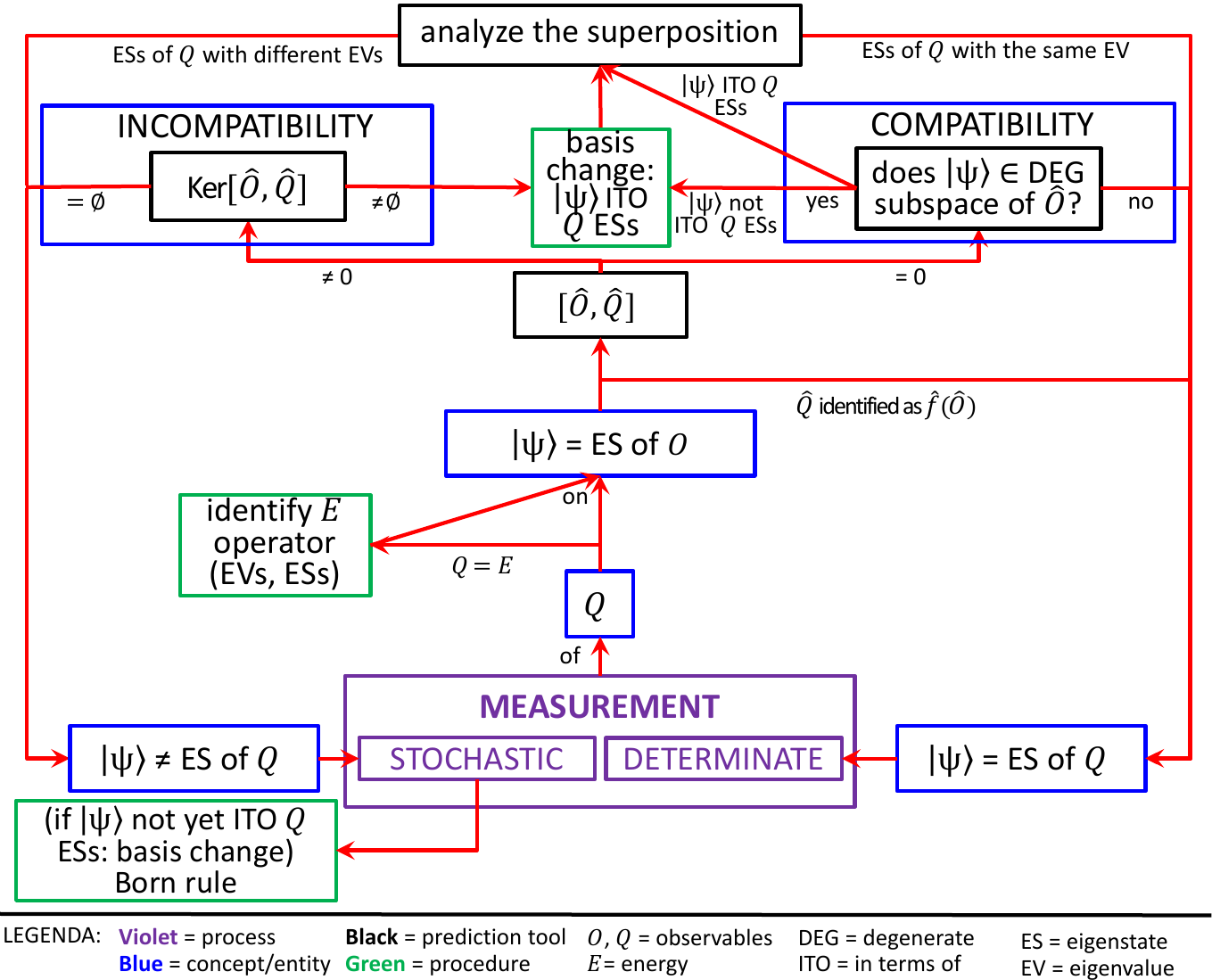}}
    \caption{Measurement on an eigenstate of an observable.}\label{FIG:7}
\end{figure}

In order to highlight how the structures encoded in the relations between one observable and another affect the outcome of the processes, we developed two additional maps. Here the state is not represented as a superposition state, but as an eigenstate of one observable. This concept represents the junction between the vector structure of the states and the operator structure of the observables, as it is a state vector, but at the same time an element of an eigenbasis of a Hermitian operator. The tool for predicting the outcome is the analysis of the commutation relations, thus establishing a structural analogy with the map on the constants of motion. As concerns the pathways of solutions on measurement and the time evolution of the state, their similarities and differences are the same as before. Therefore, for reasons of space, we include only the map on measurement (fig. \ref{FIG:7}).
The picture here is much more complex, as the relations between observables, namely compatibility and incompatibility, reveal internal structures and challenges, and the analysis of the commutator may not be sufficient for finding the results. In many cases, we need to determine whether there is degeneracy or incompatible observables with simultaneous eigenstates (the kernel of their commutator is non-empty\footnote{The kernel of an operator is the subspace of all vectors that are mapped to zero.}), and if yes, we may have to resort again to the change of basis and the analysis of the superposition. In fig. \ref{FIG:7}, we also consider the possibility of a (one-valued) functional dependency of $\hat{Q}$ on $\hat{O}$. Since the operators can be written in matrix notation, where this dependency is not directly visible on inspection, we used the expression ``$Q$ identified as $\hat{f}(\hat{O})$''. Overall, from an educational point of view, these maps are as important as those based on a superposition state, if not more so, as they allow us to explicitly visualize the fundamental issues involved in working with the concept of eigenstate and the commutation relations between observables. Using these maps requires a knowledge of a formal representation of the eigenstate concerned, while the context emerges if energy, change of basis or the examination of the degeneracy are involved.

\begin{figure}\centering
\begin{tabular}{|r|l|} \hline
    \includegraphics[width=.58\textwidth]{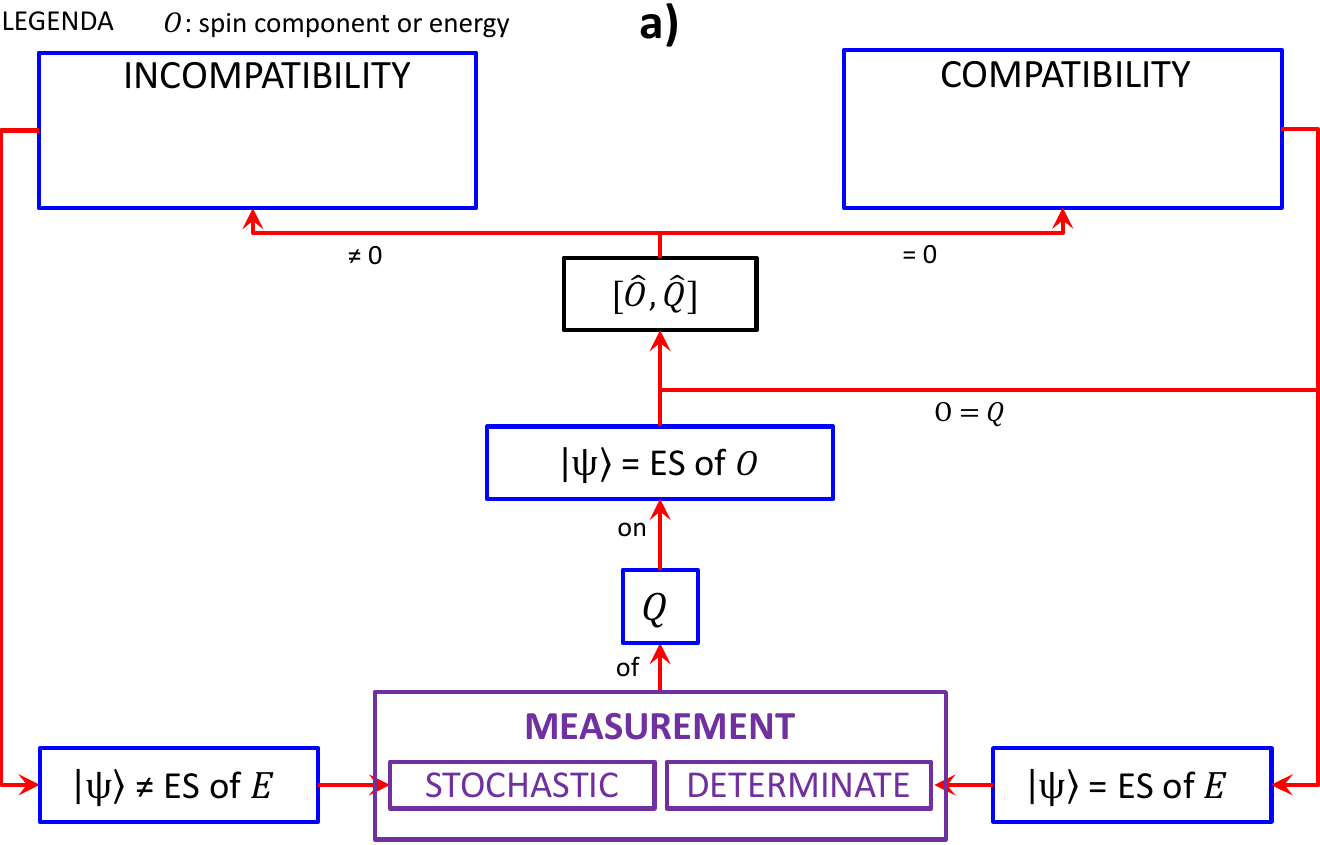} &
    \includegraphics[width=.42\textwidth]{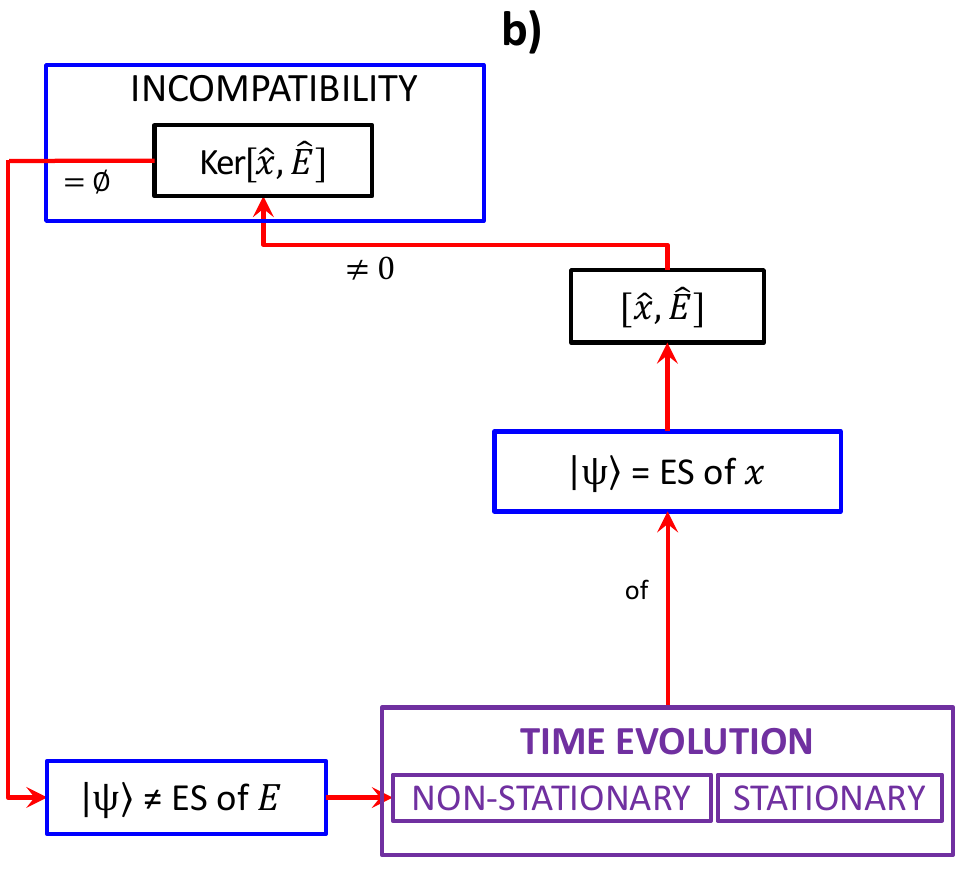} \\ \hline
\end{tabular}
\caption{a) Spin-$\frac{1}{2}$ particle in a magnetic field: measurement of the spin component $Q$ on an eigenstate of the observable $O$; b) Acontextual: time evolution of a position eigenstate.}\label{FIG:6}
\end{figure}

The new qualitative pathways of solution can be illustrated by means of two examples: one in the context of spin $1/2$, the other concerning systems in one spatial dimension (see fig. \ref{FIG:6}).

\begin{figure}[!ht]
    \centering
       \fbox{\includegraphics[width=15cm]{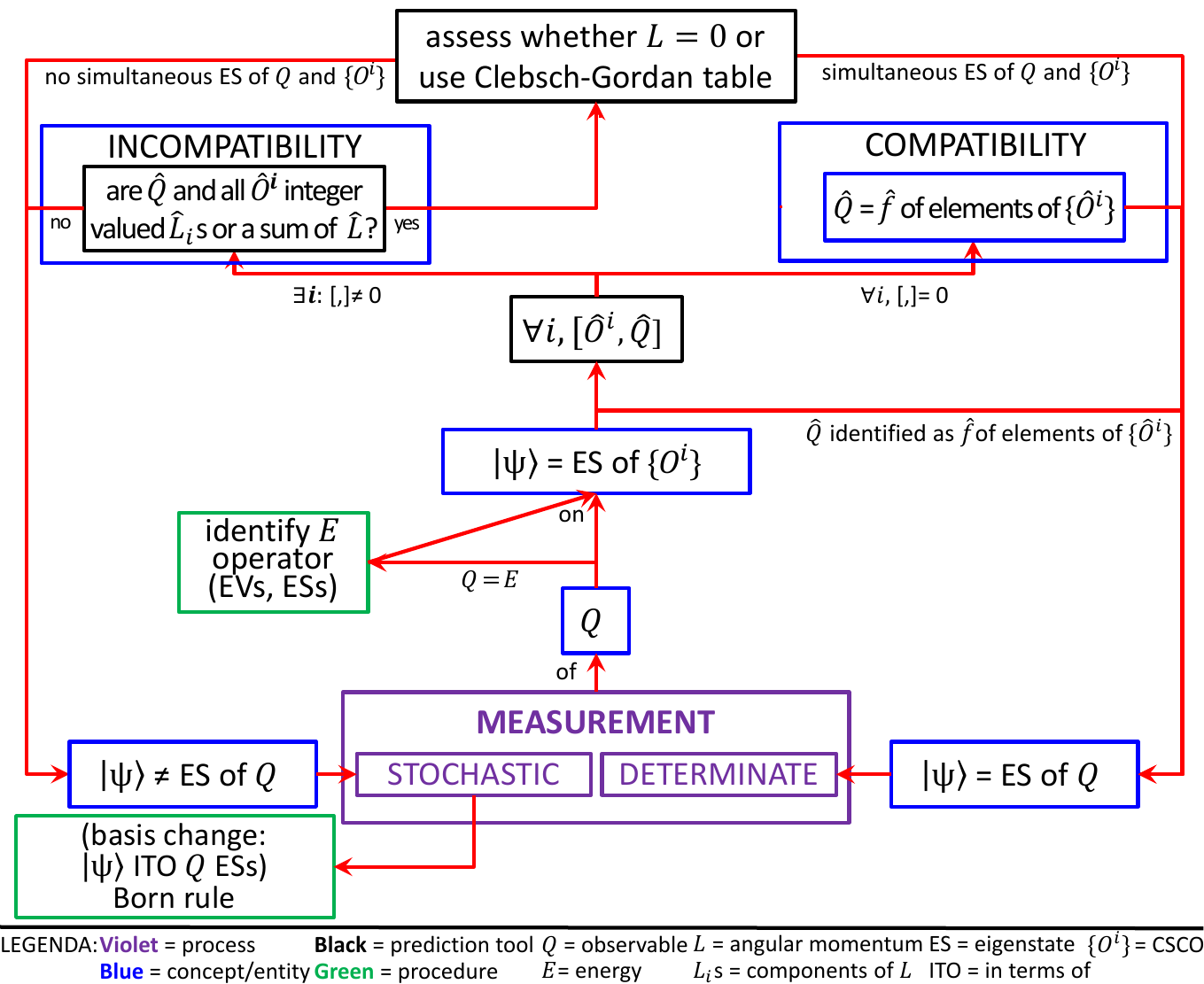}}
    \caption{Measurement on an eigenstate of a CSCO.}\label{FIG:8}
\end{figure}

A further, and last, system of maps allows us to complete the picture of the organizing principles developed thanks to this analysis. These maps display how the issues of degeneracy and a non-empty kernel - which, abstract problems aside, may emerge in undergraduate courses only when angular momenta and their addition are involved - can be removed if, instead of focusing on an eigenstate of just one observable, we refer to the simultaneous eigenstate of a CSCO. As we can see in fig. \ref{FIG:8}, now the analysis of the superposition is out of the picture and the change of basis intervenes only for obtaining quantitative results. The general pathways of solutions are uncomplicated, with the exception of a limited number of cases concerning the components of integer-valued angular momenta (orbital or spin) and their addition.

Within this framework, the nodes at the top of the knowledge pyramid are the intertwined concepts of measurement, time evolution, state, observable, eigenstate and their superposition, relations between observables (i.e. compatibility and incompatibility) and their formal representations. The prediction tools attached to these concepts are the analysis of superposition and of the commutation relations (which may involve further tools). Ancillary but needed procedures are those represented in the green boxes: the change of basis, the identification of energy eigenstates and eigenvalues for the system under examination, the calculation of probabilities (Born rule), the time evolution operator, the generalized Ehrenfest theorem.

Moving on to discuss RQ1.2, we can further highlight the educational significance of the relations between observables, since they allow us to fully explain how information on the results of the processes is organized in the formal representations of the quantum state. If we expand a vector or a wavefunction in the common basis of a CSCO, the modulus of the probability amplitude encodes complete information on measurement of the elements of these compatible observables and their operator functions. The relative phases integrate information contained in the modulus for what concerns the measurement of observables which are incompatible with at least one element of the CSCO. As concerns time evolution, we observe that if a superposition of energy eigenstates corresponding to different eigenvalues is represented with respect to the basis of a CSCO including energy, the modulus of the probability amplitude contains information on constants of motion. Consequently, it becomes clear why only phase relations may evolve in time.

Educational research reveals that interpreting the physical meaning and the measurable effects of the relative phases is a challenge even for graduate students.\cite{Michelini2014, Emigh2017} Therefore, their structural features should be emphasized in teaching. For instance, they can be explored in the simple case of position and momentum, if we do not restrict the treatment to the uncertainty principle, but focus also on the expectation values: while the modulus of the wavefunction in position space provides constraints on the corresponding wavefunction in momentum space ($\Delta x\Delta p\ge\frac{\hbar}{2}$), the phase function contained in $\psi(x)=A(x)e^{i\phi(x)}$ is needed to find the expectation value of momentum:
$$\langle p\rangle=\langle\psi|\hat{p}|\psi\rangle=\int_{-\infty}^{+\infty}dx{\langle\psi|x\rangle\langle x|\hat{p}|\psi\rangle}=\int_{-\infty}^{+\infty}dxA(x)e^{-i\phi(x)}\left(-i\hbar\frac{dA(x)e^{i\phi(x)}}{dx}\right)$$
Carrying out the integral, we obtain
$$\langle p\rangle=\hbar\int_{-\infty}^{+\infty}dx{A(x)}^2\phi^{'}(x)$$.

\section{Implementation in teaching and research}
The maps have been designed as visual tools available to instructors for helping students develop a global knowledge structure on the processes which can be applied in every context. What happens to the systems in the measurement process is not included in the maps. Therefore, these tools can be adapted to different approaches and interpretive stances.

We suggest three different ways to use the maps in teaching.

First, the maps could be used by the lecturer. Each time students are presented a new Hamiltonian, they face various changes in structures (e.g., a different Hilbert space, the appearance/disappearance of degeneracy) and procedures (the form of the eigenvalue problem, different transformation matrices). These changes particularly affect the pathways of solution represented on the maps based on the relations between two observables (see fig. \ref{FIG:7}, \ref{FIG:6}). After explaining the topic, the lecturer could show the maps to the class and walk the students through them, in order to connect each structure and procedure with the general knowledge needed to make predictions on the processes, putting the pieces together to build a global picture across different contexts.

Another way to use these tools is during recitation sections. The instructor could build the maps while solving problems, thus tracing the individual pathways of solution back to a general knowledge structure.

Finally, students could use the maps on their own, as a support while solving homework problems.


As concerns educational research, the maps may serve as a basis for the design of investigations into student understanding: the maps on the relations between two observables have been used to develop a survey on incompatibility and compatibility.\cite{Michelini2014}

\section{Indications on the organization of subject-matter content}

Educational research has established that a science curriculum for advanced study that aims to promote learning with understanding should link ``new knowledge to what is already known by presenting concepts in a conceptually and logically sequenced order that builds upon previous learning within and across grade levels.''\cite{Gollub2002} As concerns the link to the content of previous courses, such as modern physics and electromagnetism, starting with waves could be seen as a natural choice. On the other hand, a spin-first approach can be easily linked to linear algebra courses, and it can be argued that to keep focusing on waves could leave students with the impression that doing theoretical physics is just learning more and more clever tricks to solve differential equations.\cite{Harshman2019} But what about building upon previous learning within the QM course? Our analysis offers a framework for examining this issue at a global level. In fact, the aforementioned guideline can be translated into a gradual construction of the entities and the interconnections displayed in the maps, adding pieces to the puzzle until we get a complete picture.

\begin{figure}[!ht]
    \centering
       \fbox{\includegraphics[width=15cm]{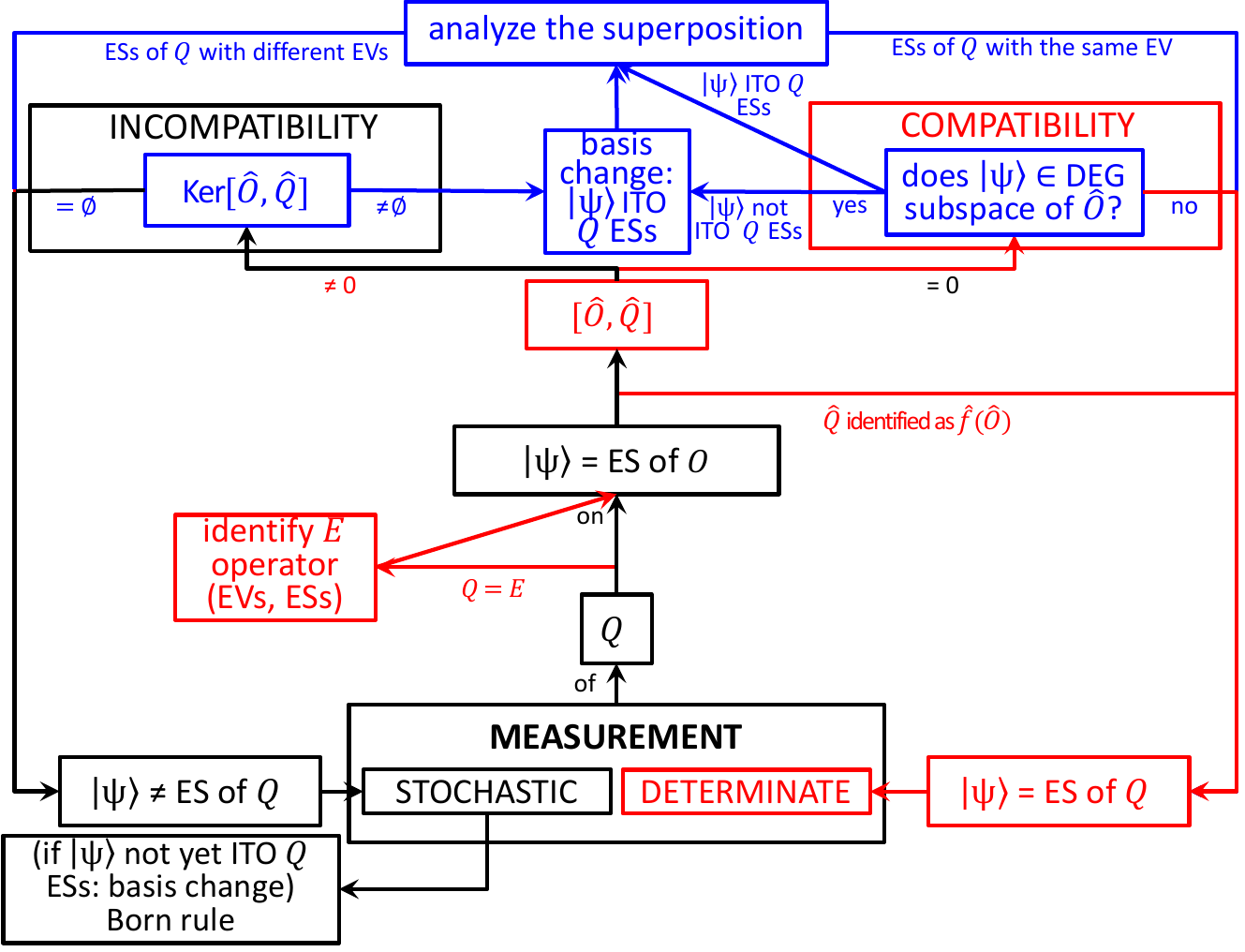}}
    \caption{Measurement: spin-first approach in steps. Step 1: black; step 2: red; step 3: blue.} \label{FIG:9}
\end{figure}

According to this framework, a spin-first approach is more suitable to the purpose. In the context of spin-1/2, the existence of an interplay between states and observables is immediately visible, as well as the meaning of a change of basis, and - more importantly - the multifaceted features of this interplay can be discovered in a very gradual way. We show it by describing how a lecturer could use the maps in teaching. In steps:
\begin{enumerate}
    \item  The course starts with measurement: every spin state is an eigenstate of a spin component; all spin components are incompatible with each other. After covering the content, we show a map based on the relations between two observables (fig. \ref{FIG:9}, step 1) with the boxes on ``incompatibility'' and ``change of basis + Born rule'', but no ``compatibility'', ``degeneracy'', ``kernel of the commutator'' - which is necessarily empty in a two dimensional Hilbert space. The organizing principles are in their simplest form. We can also add the map based on a superposition state, but only in order to discuss the superposition of two vectors when we do not know what eigenstate we are dealing with;
    \item A new observable, energy, comes into play in the context of Larmor precession. For what concerns measurement, we only need to add the box on ``compatibility'' (but no degeneracy) and possibly a new procedure: ``the identification of energy eigenstates and eigenvalues'' (fig. \ref{FIG:9}, step 2). The map concerning time evolution is presented, thus introducing a small number of new concepts, but without altering the structure of the pathways;
    \item Next, we could proceed as Townsend, by covering two-particle systems with spin-half. At this step, we complete the construction of the maps based on the relations between two observables, as we must deal with degeneracy and the possibility of a non-empty kernel of the commutator (fig. \ref{FIG:9}, step 3). At the end, we can introduce the concept of CSCO and add the maps based on it, thus providing a general solution for measurement and time evolution;
    \item The last step is the passage to QM in one and more spatial dimensions. In this context, we face many situations where it is not suitable to interpret a superposition as an eigenstate of a given observable. Therefore, we complete the exploration of the maps based on a superposition state.
\end{enumerate}
In a waves-first approach, we should start almost immediately with fig. \ref{FIG:9}, step 3, since degeneracy appears already in the context of the free particle, and nothing ensures that the kernel of the commutator of two operators (on an infinite-dimensional space) is empty.

\begin{acknowledgments}

We gratefully acknowledge Prof. Marisa Michelini and Prof. Sebastiano Sonego for their precious support and advice.

\end{acknowledgments}

\end{document}